\documentclass[prc,preprintnumbers,amsmath,superscriptaddress,amssymb]{revtex4}

\usepackage{subfigure,dcolumn}
\usepackage[T2A,T1]{fontenc}
\usepackage[russian,english]{babel}

\usepackage{listings}
\usepackage{graphicx}% Include figure files
\usepackage{subfigure}

\usepackage{lineno}

\usepackage{hyperref}
\usepackage{xcolor}
\usepackage{tikz,xcolor,hyperref}
\definecolor{lime}{HTML}{A6CE39}
\DeclareRobustCommand{\orcidicon}{
	\begin{tikzpicture}
	\draw[lime, fill=lime] (0,0) 
	circle [radius=0.16] 
	node[white] {{\fontfamily{qag}\selectfont \tiny ID}};	
	\draw[white, fill=white] (-0.0625,0.095) 
	circle [radius=0.007];	
	\end{tikzpicture}
	\hspace{-2mm}}
\newcommand{\orcid}[1]{\href{https://orcid.org/#1}{\textcolor[HTML]{A6CE39}{\orcidicon}}}

\begin{document}
\title{Development of Gated Fiber Detectors for Laser-Induced Strong Electromagnetic Pulse Environments}
\thanks{
Supported by
the National Nature Science Foundation of China (grant Nos. 11875191, 11890714, 11925502, 11935001, and 11961141003)
and the Strategic Priority Research Program (grant No. CAS XDB1602).
}

\def\stu {School of Physical Science and Technology, Shanghai Tech University, Shanghai 201210, China}
\def\siap {Shanghai Institute of Applied Physics, Chinese Academy of Sciences, Shanghai 201800, China}
\def\fdu {Key Laboratory of Nuclear Physics and Ion-beam Application (MOE), Institute of Modern Physics, Fudan University, Shanghai 200433, China}
\def\siom{Shanghai Institute of Optics and Fine Mechanics, Chinese Academy of Sciences, Shanghai 201800, China}
\def\SARI{Shanghai  Advanced Research Institute, Chinese Academy of Sciences, Shanghai 201210, China}
\def\UCA{School of Physical Sciences, University of Chinese Academy of Sciences, 100049 Beijing, China}

\author{Po Hu} 
\affiliation{Shanghai Institute of Applied Physics, Chinese Academy of Sciences, Shanghai 201800, China}
\affiliation{\stu}
\affiliation{School of Nuclear Sciences and Technology, University of Chinese Academy of Sciences, Beijing 100049, China}
\affiliation{\fdu}
%\author{YYYY\normalfont\textsuperscript{*}} \affiliation{\sjtu}
%\thanks{These two authors contributed equally}
\author{Zhiguo Ma} \affiliation{\fdu}
\author{Kai Zhao} \affiliation{\fdu}
\author{Guoqiang Zhang}  \affiliation{\SARI}
\author{Deqing Fang}  \affiliation{\fdu}
\author{Baoren Wei}  \affiliation{\fdu}
\author{Changbo Fu} \email[Corresponding author:] {cbfu@fudan.edu.cn} \affiliation{\fdu}
\author{Yu-Gang Ma\orcid{0000-0002-0233-9900}} \email[Corresponding author:] {mayugang@fudan.edu.cn} 
\affiliation{\stu}
\affiliation{\fdu}

\begin{abstract}
With the development of laser technologies, 
nuclear reactions can happen  in high-temperature plasma environments induced by lasers
and have attracted a lot of attention from different physical disciplines.
However, studies on nuclear reactions in plasma are still limited by detecting technologies.
This is mainly due to the fact that 
extremely high electromagnetic pulses (EMPs) can also be induced 
when high-intensity lasers hit targets to induce plasma,
and then cause dysfunction of many types of traditional detectors.
Therefore, new particle detecting technologies are highly needed.
In this paper, we report a recently developed gated fiber detector 
which can be used in harsh EMP environments.
In this prototype detector, scintillating photons are coupled by fiber
and then transferred to a gated photomultiplier tube  
which is located far away from the EMP source and shielded well. 
With those measures, the EMPs can be avoided, 
and this device has the capability to identify a single event of nuclear reaction products generated in laser-induced plasma from noise EMP backgrounds.
This new type of detector can be widely used as a Time-of-Flight (TOF) detector in high-intensity laser nuclear physics experiments for detecting neutron, photons, and other charged particles.
\end{abstract}

\keywords{Gated Fiber Detector; Radiation Detection; High-Intensity Laser; Strong Electromagnetic pulses}

\maketitle

%\usepackage{CJK}

%\def\etal {{\it et al.}}
%\def\dd {{\rm d}}

%\usepackage{makeidx}
%\nofiles
%\makeindex

%\begin{document}

%\linenumbers

%\preprint{}

%\altaffiliation[also at]{}

%\pacs{25.70.Ef, 29.27.-a}% PACS, the Physics and Astronomy

%\tableofcontents
%=================================================================
%\maketitle

\section{Introduction}

With the development of high-intensity laser (HIL) technologies, 
it is possible today to create plasma environments for fundamental nuclear studies or nuclear applications
\cite{ditmire1997high,
Laser-Acc-tajima1979laser,
FU20151211,
n012-laser-klir2015efficient,
ELI-Balabanski-2017,
DD-zhang2017deuteron,
Laser60TW-p-CPC2018,
LuoWen-Cu62-2019Photonuclear,
Luowei-2018QED,
2017Transmutation}.
For example, 
in inertial confinement fusion experiments, 
multiple ns-pulse-width HILs can compress targets 
which are composed of deuterium or tritium  
to densities as high as 10,000 times of theirs initial\cite{RN3}, 
and then igniting nuclear reactions. 
High energy neutrons and protons can be produced in this process through reactions
${\rm D(D,n)^3He}$, ${\rm D(D,p)^3H}$, or ${\rm D(T,n)^4He}$, etc.
In laser Coulomb explosion experiments\cite{PhysRevLett.84.2634,TORRISI201342}, 
a fs-pulse-width HIL hits on deuterium nano-clusters, 
strips their electrons,
and then cause coulomb explosion of the deuterium ions. 
Nuclear reactions are triggered when D ions colliding with each other.
Nuclear reactions have also been observed  
in so-called ``laser plasma collider'' scheme\cite{1999Nuclear,FU20151211,DD-zhang2017deuteron},
where plasmas induced by ns-pulse-width HILs collide with each other head-on-head.
With more and more HIL facilities running or under construction, 
this new interdisciplinary, so-called laser nuclear physics,  will have a brilliant future. 

However, methods of detecting nuclear products induced by HILs are still limited 
and needed urgently. 
In over 100 years  of nuclear and particles physics history, 
various types of detectors have been developed for different environments, 
such as
scintillating photon based detectors (for examples, plastic, liquid, and gas scintillator), 
semiconductor based detectors (for examples, Si, high purity germanium, and diamond, etc.),
and traced detectors (for examples, CR39 etc.)\cite{DetectorTech.book.2010, CR39-2019-GSL-article}.
However, most of these detecting technologies cannot be used directly in HIL environments because of the following difficulties. 

In a typical high laser experimental environment, 
electromagnetic pulses (EMPs)\cite{Laser-Ablation-CPC2020, Collisionless-Shockwave-He_2019,Laser60TW-p-CPC2018} can interfere with many types of traditional detectors,
and cause them dysfunction. 
When a high intensity laser focuses on a target, 
it interacts with the target's materials 
and causes emitting of  photons (or in other words, electromagnetic waves) almost at any frequency.
The photon spectra could cover radio frequency, microwave,  infrared, optical,  
X-rays, as well as  $\gamma$-ray domains. 
For semiconductors, bias voltages are needed when running them.
However, in high EMP environments, electromagnetic fields of EMP can be much larger than 
the applied bias voltages (fields) and then cause errors or even damage to the detectors.
The same problems appear in photomultiplier tubes (PMTs) too.
EMP fields could  distort the fields applied between a PMT's dynodes highly 
and then causes its dysfunction.
Therefore, even scintillators themselves may also work under EMPs, 
but due to the dysfunction of PMTs coupled to them, 
traditional setups still cannot work in strong EMP environments.
EMPs can also cause dysfunction of electronics, including amplifiers, amplitude-to-digital converters (ADCs), and computers, etc.
Particularly, strong microwaves generated during laser-target interactions 
are recognized as a threat to electronics and computers\cite{2020Laser}. 

Trace detectors like CR39\cite{CR39-nature2017-RN98,CR39-ZhangYue-2019article, CR39-ZhangYue-2020article,Bulk-CR39-2020article} 
or Thomson spectrometers which record particles with image plates\cite{ThompsonSpec2019-CPC,TomSpc2020.RN99} etc. 
are not sensitive to the EMPs.
Therefore, they have been widely used in HIL experiments today. 
However, their detecting sensitivities are very limited, 
and they are not so convenient to be used too.
Detecting nuclear reaction products down to a single particle in HIL environments is still challenging today.
Developments of new robust radiation detectors are very important for further progress in laser nuclear physics.

In this paper, we present a recently developed gated fiber detector (GFD) which can be used for strong electromagnetic environments\cite{FiberDet-Patent2020}. 
In the second section, the structure of the GFD will be described,
and in the third section, online testing results will be given,
followed by a summary.

\section{Structure of the Gated Fiber detector}\label{sec.II}
\begin{figure*}[!htb]
\includegraphics[width=\hsize]{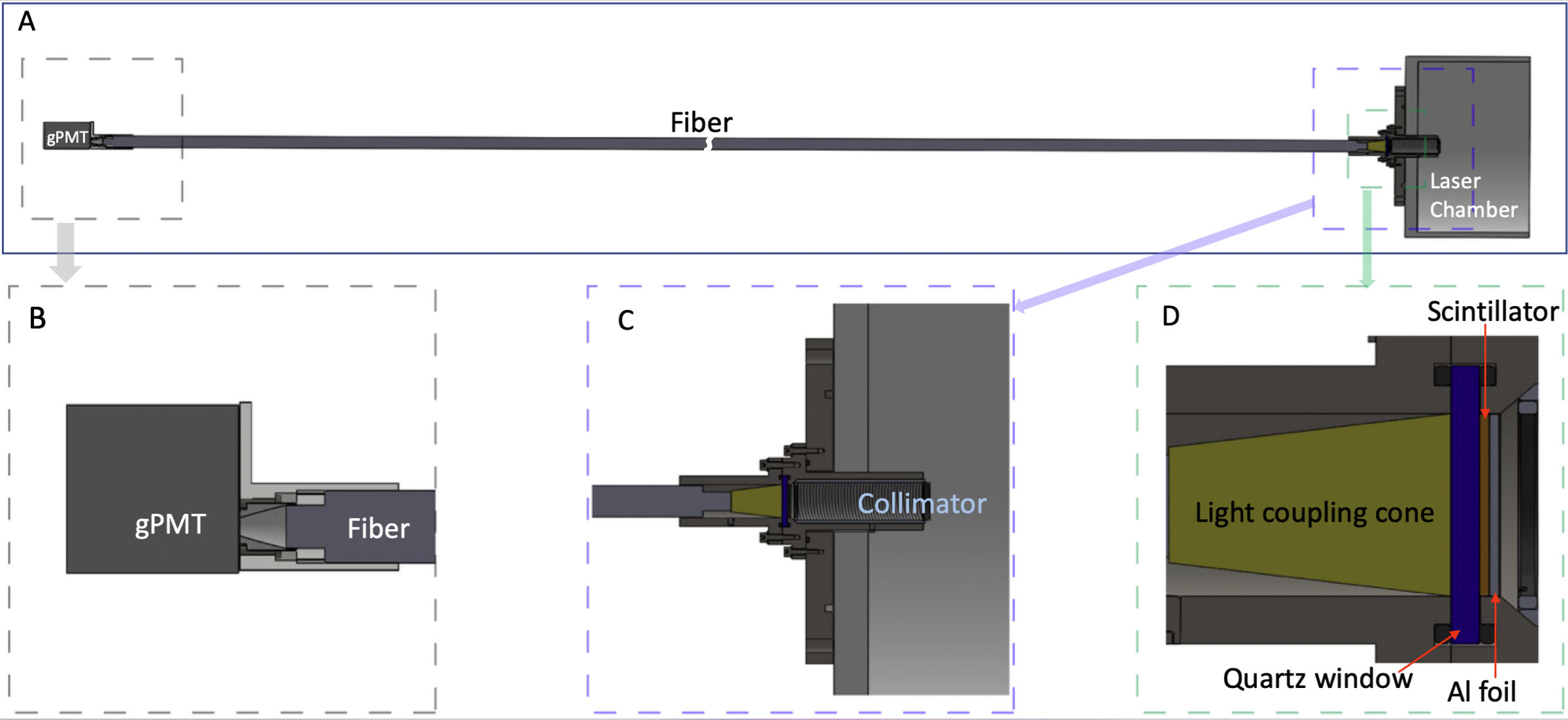} 
\caption{(Color online) Schematic drawing of  a gated fiber detector (GFD). 
(A) is the full view of the GFD, and (B)--(D) are zoom-in structures of different parts.
As shown in (C) and (D), it has a reflecting Al foil layer, a scintillating layer, a quartz glass window for sealing vacuum, and a photon coupling cone for collecting scintillator photons. 
The length of the fiber is adjustable, which can transport scintillating photons to a distance wanted, and then avoid strong EMPs in the target area.
The gPMT can be turned off at the moment of EMPs arriving to further reduce EMP effects.
}
\label{fig.OverallSetup}
\end{figure*}

A schematic drawing of a gated fiber detector (GFD)  is shown in Fig.\ref{fig.OverallSetup}.
The GFD has the following main parts, 
a reflective layer,
a scintillating layer,
a vacuum sealing glass window,
a photon coupling cone,
a fiber,
and a gated photomultiplier tube (gPMT).

%=============================
%=============================
\subsection{Reflective layer}
The reflective layer has two functions.
First, it reflects the scintillating light,
and then highly improves the light collecting efficiency. 
Second, it can reflect the original laser to protect electronics and PMT followed.
Wavelengths of HILs today are normally in the range 200-2000 nm 
with intensities up to $10^{22}$ W/cm$^2$\cite{PhysRevSTAB.5.031301},
compared with a typical PMT's working scintillating light intensity of
$10^{-9}$ W/cm$^2$, or in other words,  detecting a single photon.
If a tiny small amount of the original photons from the main laser enter into the PMT, 
it can cause the PMT to be blind which may need a long time to recover
or make the PMT totally damaged in the worst case.
Even after diffuse reflecting inside a laser target chamber,
the straggling light intensity may be reduced down several orders of magnitude.
However, it is still too strong for a PMT to accept directly, and can kill the PMT easily.

A thin layer of aluminum, Polytetrafluoroethylene (PTFE), or barium sulfate(BaSO$_4$), etc. can be used for this purpose\cite{reflector2008-article,Reflector2012-article,LYSO-BaSO42018,BaSO4-Schutt:1974}. 
With a few $\mu$m Al, charged particles can relatively easily pass through. 
For light with a wavelength in the range of $>200$ nm,
a 20 $\mu$m layer of Al can reduce the original incoming laser intensity to more than $10^{-6}$ times smaller.  
At the same time, 
this layer can reflect the scintillating light with an efficiency of about 80\%-99\%\cite{reflector2008-article,LYSO-BaSO42018},
which means that the scintillating light collecting efficiency can be almost doubled.

Al can be oxidized relatively easily if exposed to the atmosphere.
The oxidized layer (Al$_2$O$_3$) could be as thick as a few $\mu$m. 
The oxidization will not cause problems here due to the facts following.
On one hand, the oxidized layer can also stop the original laser.
On the other hand, the surface towards the scintillator is airtight, and this side cannot be oxidized.
It still can serve as a mirror to reflect the scintillating light lasting for a very long time.

A list of reflectors which are frequently used in scintillators is shown in Tab.\ref{tab.reflector}. 
As shown in the table, a reflecting coefficient of 0.99 is achievable.
\begin{table}[]
\caption{A list of reflectors commonly used for scintillating experiments.}
\begin{tabular}{cccc}
\hline \hline
Reflector & Refl.Coeff.(@440nm) &thickness(mm)&Ref.\\ 
\hline
PTFE tape&0.99&0.06&\cite{reflector2008-article,Reflector2012-article}\\
Magnesium Oxide&0.98&1&\cite{BaSO4-1968-Grum:68}\\
Barium Sulfate&0.98&0.12&\cite{BaSO4-1968-Grum:68,BaSO4-Schutt:1974}\\
Titanium dioxide paint&0.95&0.14&\cite{Reflector2012-article}\\
Aluminum Foil&0.79&0.025&\cite{reflector2008-article}\\
Tyvek paper&0.98&0.11&\cite{reflector2008-article}\\
\hline \hline
\end{tabular}
\label{tab.reflector}
\end{table}

%=============================
%=============================
\subsection{ Scintillator layer}
For different physical purposes, one can choose different materials for the scintillating layer.
For HIL applications, normally a fast rising-time is required.
Therefore, we prefer fast response scintillators like plastic, 
as well as inorganic ones including  
LYSO(${\rm (LuY)_{2}SiO_5:Ce}$),
YAP(YAlO$_3$), 
YAG(${\rm Y_3Al_5O_3}$),
and LSO(${\rm Lu_{2}SiO_5:Ce}$) etc\cite{LYSO-BaSO42018}.
The properties of them are listed in Tab.\ref{tab.scin.properity}.

If there has a high neutron flux in the background, 
inorganic scintillates are preferred.
Organic scintillates normally have a high percentage of hydrogen inside.
Neutrons  have a very high scattering cross section on hydrogen which results in high background noise.

If using GFD to detect charged particles, a vacuum is necessary. 
Therefore, after the scintillate layer, 
a quartz glass window is employed to separate the target chamber vacuum from the atmosphere.

Hygroscopic scintillating materials are not convenient to be used.
They have to be sealed totally to avoid them catching moisture in the air.
If for detecting charged particles, 
the sealing materials would be a dead layer
and then cause an extra measurement uncertainty.
Therefore, non- or low- hygroscopic scintillators are preferred.
The hygroscopic property of different scintillators is also listed in the Tab.\ref{tab.scin.properity}.

\begin{table*}[]
\caption{Comparison of major parameters of different type scintillators which have relatively fast light decay times.}
\begin{tabular}{cccccccccc}
\hline \hline
Properties & LYSO &LSO& plastic(EJ200)  &YAP:Ce&YAG:Ce&PWO&CsI&LaBr$_3$:Ce\\ 
\hline
Density(g/cm$^2$)  & 7.1&7.35 & 4.51 &5.37&4.57&8.3&4.51&5.29\\
photon yield[photons/MeV]&33200& 30000&10000&10000&8000&120/40 & 34000/18000 &52000\\
Decay time(ns)&36&40&2.1&25& 70&30/10&30/6&20\\
Peak Wavelength(nm)&420&420&425&370&550&425/420&420/310&356 \\
Refractive index&1.81&1.82&1.58&1.95&1.82&2.20 &1.79&1.9\\
Hygroscopic &No&No&No &No&No&No&Slight &Yes\\
\hline \hline
\end{tabular}
\label{tab.scin.properity}
\end{table*}

%=============================
%=============================
\subsection{Fiber coupling}
In a typical HIL experiment,
EMPs are very strong near the targets.  
With distance to the target increases, the EMP will be weaker and weaker, 
roughly following inverse square law. 
The closer the detectors and electronics to the target,
the stronger the EMP they will suffer.
Furthermore, because of the limited space, 
electronics cannot be shielded fully to avoid impacts from EMPs.
Therefore, by using optical fiber to transfer the scintillating light to a faraway location from the target, 
EMPs can be reduced,
together with better shielding for electronics with more materials. 
One expects a much smaller background noise by using fiber.

When choosing an optical fiber, 
the main factors are its transmission attenuation at a different wavelength and its numerical aperture (NA). 
As shown in Tab.\ref{tab.scin.properity},
wavelengths of scintillators for general purposes are in UV region of about 300--500 nm.
Therefore, UV fibers made from quartz or liquid can be used.
For a fiber which has a specific NA, 
only scintillating light which has an incident angle $\theta$\cite{Fiber-coupling-Elsey:07},
\begin{equation}\label{eq.theta.max}
\theta \leq ArcSin({\rm NA}\cdot\frac{n_1}{n_2})\equiv\theta_{max},
\end{equation}  
can pass through it,
where $n_1/n_2$ is the refractive index of the light cone/fiber.

For quartz fiber, the larger the diameter, the harder it to be bent. 
Therefore, quartz fibers with diameters larger than 1.5 mm are hardly founded in markets. 
For  liquid UV fibers, they can be made with diameters over 10 mm.
However, transmission attenuation of liquid UV fibers is normally larger than quartz UV fibers.
A typical liquid UV fiber has an attenuation of 0.4 dB/m at 400 nm, 
compared with that of quartz, 0.05 dB/m.

%=============================
%=============================
\subsection{Light coupling cone}

It can be proved that lens coupling cannot have a higher efficiency than the end-to-end coupling method.
In fact, if using a lens to focus scintillating light onto the ends of a fiber,
even the light intensity on the fiber's ends increases, 
the angle spreading increases at the same time.
Therefore, the light collecting efficiency does not increase at all,
because fibers can only accept light with an incident angle 
smaller than $\theta_{max}$ in Eq.\ref{eq.theta.max},
Therefore, we design a light coupling cone, and not a lens, for improving the light collecting efficiency.

Materials like quartz or Polymethylmethacrylate (PMMA) can be used to make the cones 
for their low attenuation in the ultraviolet range.
The structure of the light coupling cone is shown in Fig. \ref{fig.OverallSetup}(D).
At the scintillator end, 
because of the reflective layer which is described in the previous subsection,
the light collecting efficiency is almost doubled.
The cone's side surface is painted too to reflect the scintillating light.
Painting materials like EJ-510, Al, Ag, and BaSO$_4$, etc. can be used, because of their reflect coefficients  in wavelength range 300--600 nm.

%=========================
%=========================
\subsection{Gated PMT}
%=========================
%=========================

To overcome the strong EMPs caused by original main laser pulses,
as well as other possible laser-related backgrounds, like laser-induced neutrons,
a gated PMT detector will be used.

A gPMT is a photo-multiplier tube with a gating circuit. 
The ``gate'' here is different from normal detector gates.
Normally detectors' output signals are gated. 
In that way, one can choose to use or not use an output signal, 
but the detector itself is always on and working.
If very strong EMPs coming, only gating outputs does not help. 
While here it is designed to be that 
the detector itself, 
specifically the bias voltage of the PMT's dynodes, 
are gated to be power on or power off.
Once an EMP coming, 
the bias of the PMT is turned off,
and this will protect the detector as well as the following electronics 
not to be affected or even damaged by the EMP.

\begin{figure}[!htb]
\subfigure{
\includegraphics[width=8.5cm]{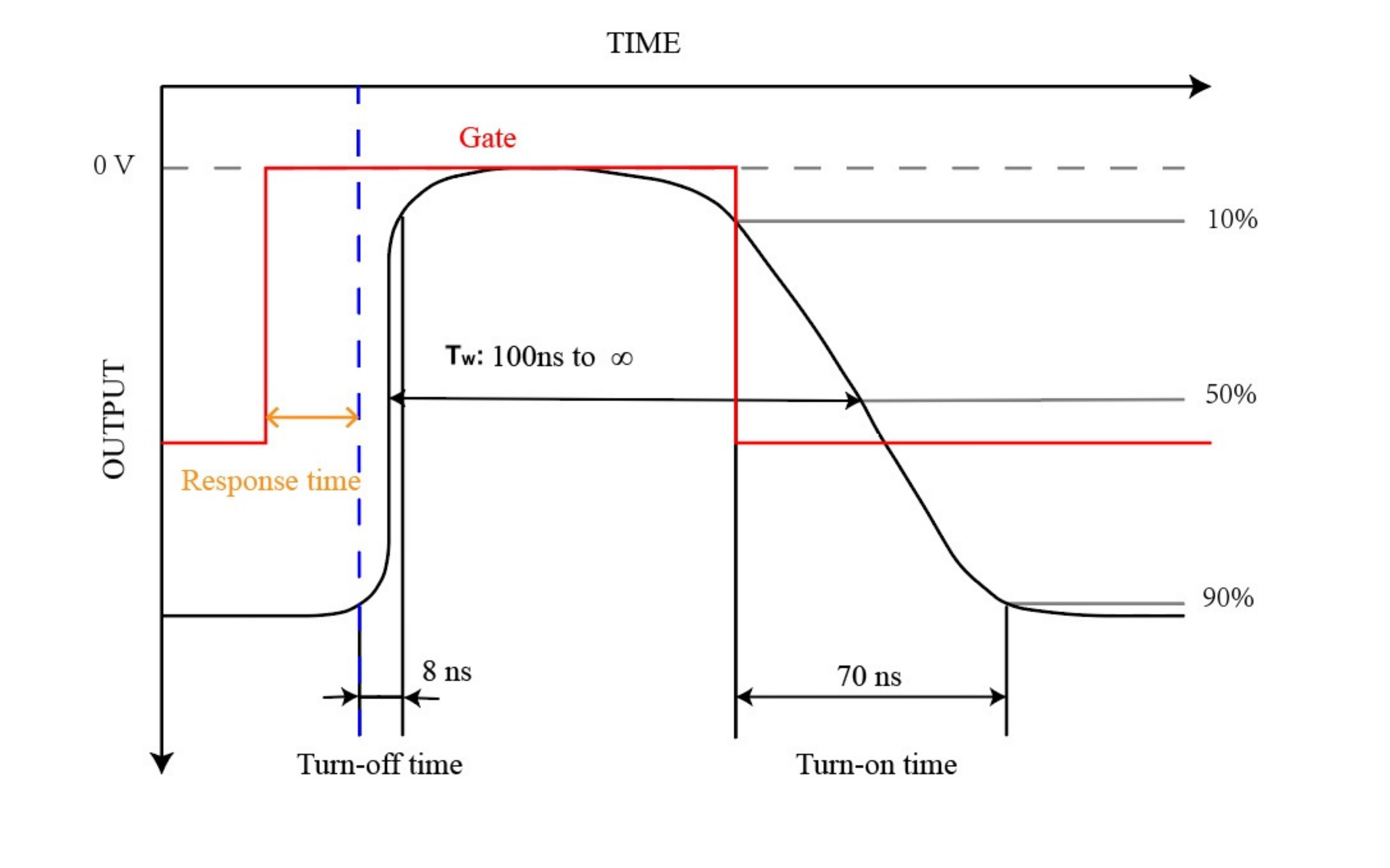}
}
\caption{
(Color online) A schematic of the timing sequence of the gPMT which works in the normal-on model.
This gPMT can respond to a gate signal quickly. 
Specifically, it can turn off the power supply to the PMT in 8 ns, and turn it on in 70 ns.
The time window $T_w$ can be tuned in the range of $100\ ns<T_w<\infty $ by the width of the gate signal.
In this normal-on model, the PMT will be shut down in the time window $T_w$.
}
\label{fig.gPMT-Time}
\end{figure}

As shown in Fig.\ref{fig.gPMT-Time}, we used a circuit that can close the PMT's bias voltage in 8 ns, 
and turn it on to a working condition in 70 ns. 
The time window keeping detector working, $T_{w}$, is $100\ ns<T_{w}<\infty$.
In typical HIL experiments, 
EMPs may last from a few ns to tens of ns.
Therefore, the gated PMT here can provide protection for detecting electronics.

\begin{figure}[!htb]
\subfigure{
\includegraphics[width=8.5cm]{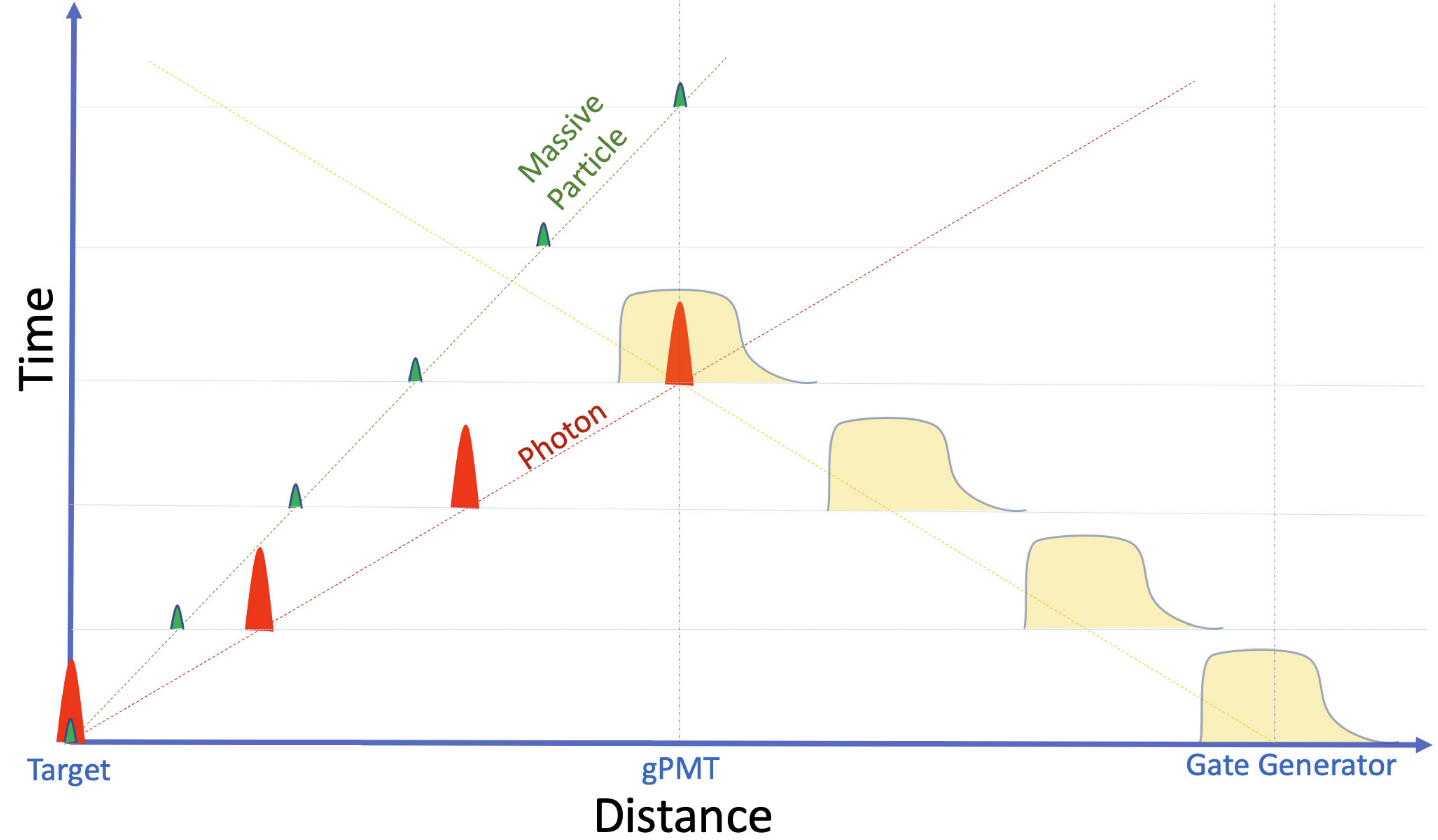}
}
\caption{
(Color online) A schematic space-time drawing of photons and massive particles traveling after lasers bombarding a target.
At time t=0, both photons and massive particles are generated at the same time.
The photons move with a speed of c (the red line), while the massive particles a lower speed (green line).
By send a gate signal (the yellow one) to the gPMT, photon signals from EMP can be vetoed, while the massive particle signals can be recorded by the gPMT.
}
\label{fig.gPMT.timing}
\end{figure}

Time evolution of EMPs and massive particle signals to be detected  are shown in Fig.\ref{fig.gPMT.timing}.
As shown there, 
EMPs and massive reaction products to be detected are generated at time T=0.
Because the massive particles have a slower speed than the EMPs,
they arrive at gPMT position earlier than the massive particles.
Therefore, by sending a gate signal from a gate generator,
one can make the gPMT not respond to the EMP signals, 
but respond to the massive particle signals.

The massive particles here could be neutrons or charged particles.
In fact, besides massive particles generated at around T=0,
any particles, including photons, which have different TOF, 
can be detected in this setup.
For example, photons that are emitted from excited nuclei, i.e. nuclear isomers,  have a longer TOF than the original EMP.
Therefore, this kind of photons, as well as other massive particles which are generated at larger time T of course,  can be recorded this way.

%========================
%========================
\section{Simulation of Scintillating Photon Collection}
\begin{figure}[!htb]
\subfigure{
\includegraphics[width=8cm]{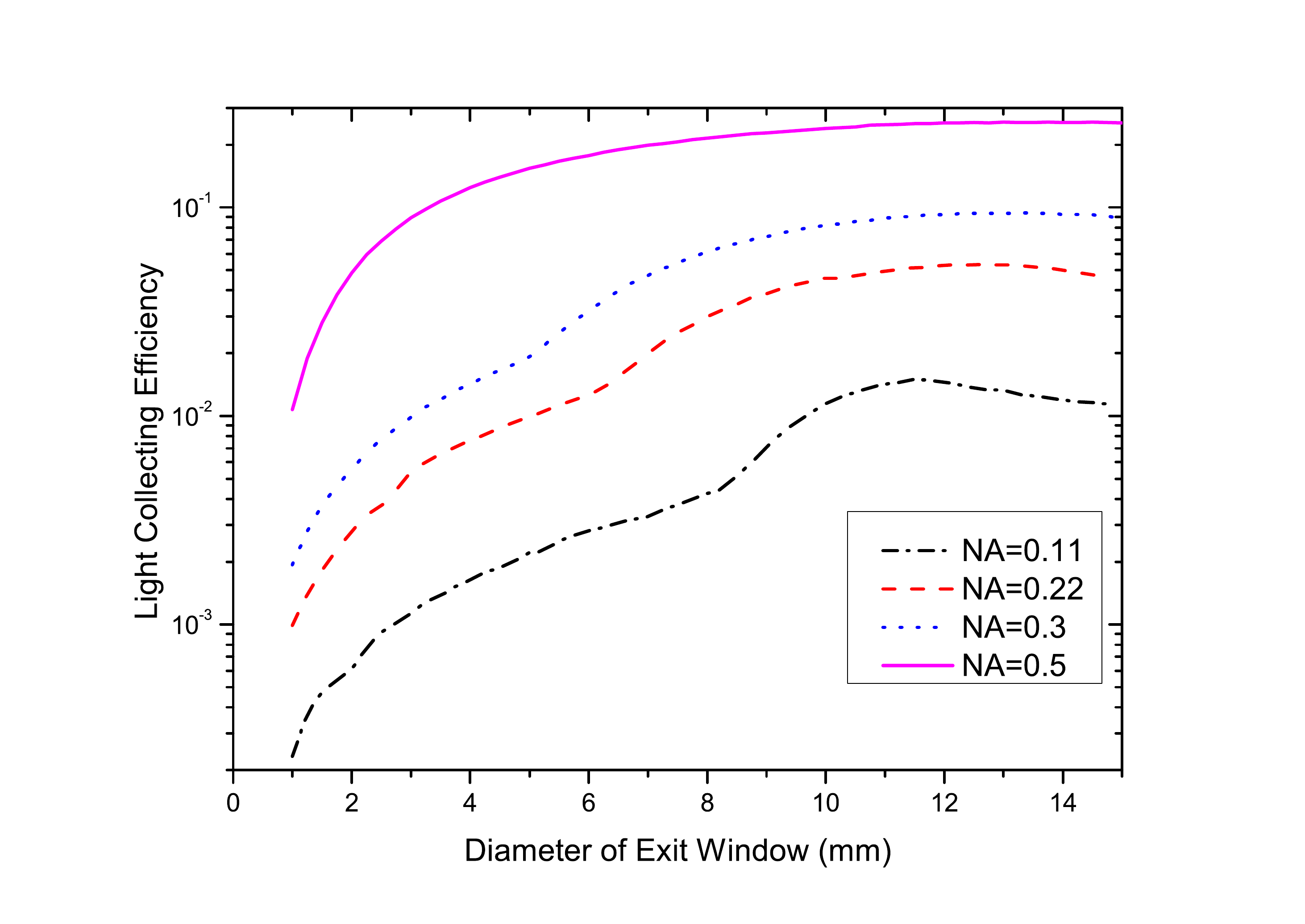} 
}

\caption{
(Color online) Simulation results of the light collecting efficiency by a cone 
which has an entrance window diameter of  10 mm, a height of 8 mm, and a variable diameter of exit window.
The cone is coupled to fibers with different NAs of 0.11, 0.22, 0.3, and 0.5.
The reflective efficiency is set to be 98\%.
}
\label{fig.cone.eff}
\end{figure}

%========================
%========================
A numerical simulation is carried out to optimize the light collecting efficiency of the cone.
A home-written programmer based on the ROOT is used. 
From the scintillator, a random scintillating ray was generated, 
and transfers inside the light-collecting cone which has a diffused surface. 
At $i$-time the ray hits the cone's surface or the reflective mirror layer, 
the corresponding reflection coefficient $R_i$ is recorded. 
Once the ray hit the exit window, depending on the angle $\theta$ between the ray and the surface of the exit window, the value $I=I_0\prod_i R_i$ will be recorded as the light intensity which passes through the fiber if $\theta\le\sin(NA)$, or discarded if $\theta>\sin( NA)$, 
where $I_0$ is the intensity of the original scintillating ray.
The followings simulation inputs parameters are  assumed: 
reflection coefficient of the side painting $R_{p}=98\%$; 
reflection coefficient of the reflective mirror layer $R_{m}=98\%$;
the scintillator diameter (entrance window) $D_s=10$ mm;
the fiber (exit window) diameter $1\le d_e\le 15$ mm;
and isotropic emitting angle of the scintillating photons.
The light coupling efficiency $\eta$ as a function of $D_e$ and NA is shown in in Fig. \ref{fig.cone.eff}. 
One can see that $\eta$ is high at $D_e=8$ mm.
Only liquid fibers are available with such high $D_e$. 
Quartz fibers with $D_e=1.5 $mm and NA=0.5 are available,  
and the coupling efficiency $\eta>1\%$ can be expected by using them.

In Tab.\ref{tab.NA}, light-collecting efficiencies for different core diameters and NAs are listed. 
As shown there, with a smaller diameter,
light collecting efficiency reduces quickly as diameter dropping.
In fact, according to the second thermodynamic law,
the percentage of the photons which has incident angle $\theta<\theta_{max}$ keeps as a constant
whatever shape of the reflection surfaces are.
Therefore, if the reflecting efficiency of the surfaces is 100\%, 
the collecting efficiency will keep as a constant too with whatever size of the existing window's diameter.
It is the reflected times $N$ that reduces the collecting efficiency dramatically.
This is because of the collecting efficiency $\eta\propto r^N$. 
Even the $r$ is close to 1,  $r=0.98$, the power $N$ makes it drops very fast.
Clearly, the smaller N, the higher the collecting efficiency $\eta\propto r^N$.
With rough surfaces and diffusing reflection, 
a ray goes randomly, 
which results,
\begin{equation}\label{eq.N.refl}
 N=\frac{A_{tot}}{A_{ex}},
\end{equation}
where $A_{tot}$ is the area of the light coupling cone's outside surface
and $A_{ex}$ is the area of the existing window.
From the Eq.\ref{eq.N.refl} one can find that $N$ could be very large!
A mirror surface together with a carefully designed geometry may be helpful. 

\begin{figure}[!htb]
\subfigure{
\includegraphics[width=8.5cm]{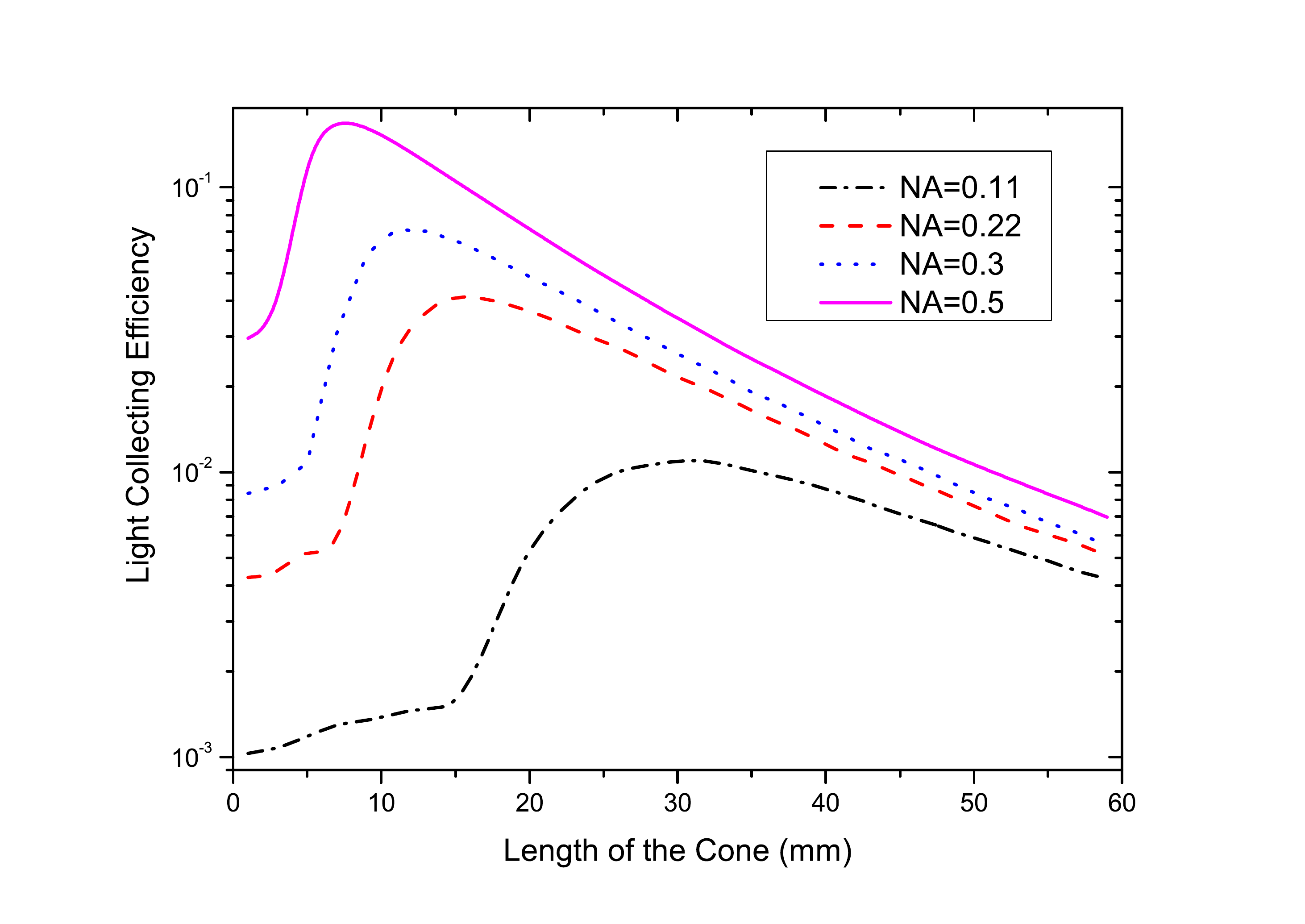} 
}

\caption{
(Color online) Simulation results of the light collecting efficiency by a cone 
which has an entrance diameter of  5 mm, an exit window diameter of 1.5 mm,
and a variable length.
The cone is coupled to fibers with different NAs of 0.11, 0.22, 0.3, and 0.5.
The reflective efficiency is set to be 98\%.
}
\label{fig.cone.hight}
\end{figure}

To optimize the light collecting efficiency $\eta$, the dependence of the cone's length $L$ is also studied as shown in Fig.\ref{fig.cone.hight}.
When  $L\to 0$, the light entrance window touches the existing window, 
and then 
\begin{equation}
\lim_{L\to 0}\eta \propto\frac{A_{ex}}{A_{in}},
\end{equation}
where $A_{in}$ is the area of entrance window.
When the $L$ increases, more and more scintillating photons have higher chance to hit the exit window with angles close to $\frac{\pi}{2}$, which results a higher $\eta$.
However, the longer the $L$, the more times of reflection. 
Because of not perfect reflection efficiency, 
more and more photons will lose during their path to the existing window, 
and then result in $\eta$ drops.
This effect can be founded in Fig.\ref{fig.cone.hight}.
For different NA fibers, there has an optimized length.
For $NA=0.5$, the $\eta$ has a peak value at around $L=6$ mm.

\begin{table}
\caption{
The relationship between NAs of fibers and light collection efficiencies.
$\theta_{max}$ is the maximum light cone angle at which light can pass through the fiber which has the corresponding NA.
$\eta_{sim}$ is the simulated light collecting efficiency, 
which has assumptions including reflecting efficiency $r=98\%$ (and perfect r=100\% also shown in brackets), 
the diameter of coupling light cone entrance window  10 mm, 
length of 8mm, and existing window diameter $D_e=1.5, 8$ mm. 
}
\begin{tabular}
{c c c c } %p{2cm}<{\centering}
\hline \hline
NA & $\theta_{max}$  
& $\eta^{(1)}_{sim}$ ($D_e=8.0$ mm) 
\footnote{Simulation inputs: r=98\% and r=100\% (in brackets).}  
& $\eta^{(2)}_{sim}$($D_e$=1.5 mm)
\footnote{Simulation inputs: r=98\% and r=100\% (in brackets).}  \\ 
\hline
0.11 & 6.3$^{\circ}$   &  $4.2\times10^{-3}$ $[4.6\times10^{-3}]$  & $4.5\times10^{-4}$ [$1.0\times10^{-3}$]\\
0.22 & 12.7$^{\circ}$ & $2.7\times10^{-2}$  $[3.1\times10^{-2}]$  & $1.8\times10^{-3}$ [$4.5\times10^{-3}$]\\
0.30 & 17.5$^{\circ}$ & $6.1\times10^{-2}$  $[6.9\times10^{-2}]$  & $3.7\times10^{-3}$ [$8.8\times10^{-3}$]\\
0.50 & 30.0$^{\circ}$& $2.2\times10^{-1}$  $[2.4\times10^{-1}]$  & $2.8\times10^{-2}$ [$7.2\times10^{-2}$]\\
\hline \hline
\end{tabular}
\label{tab.NA}
\end{table}

%========================
%========================
\section{Testing on a HIL beam line}
%========================
%========================
\begin{figure}[!htb]
\subfigure{
\includegraphics[width=8.5cm]{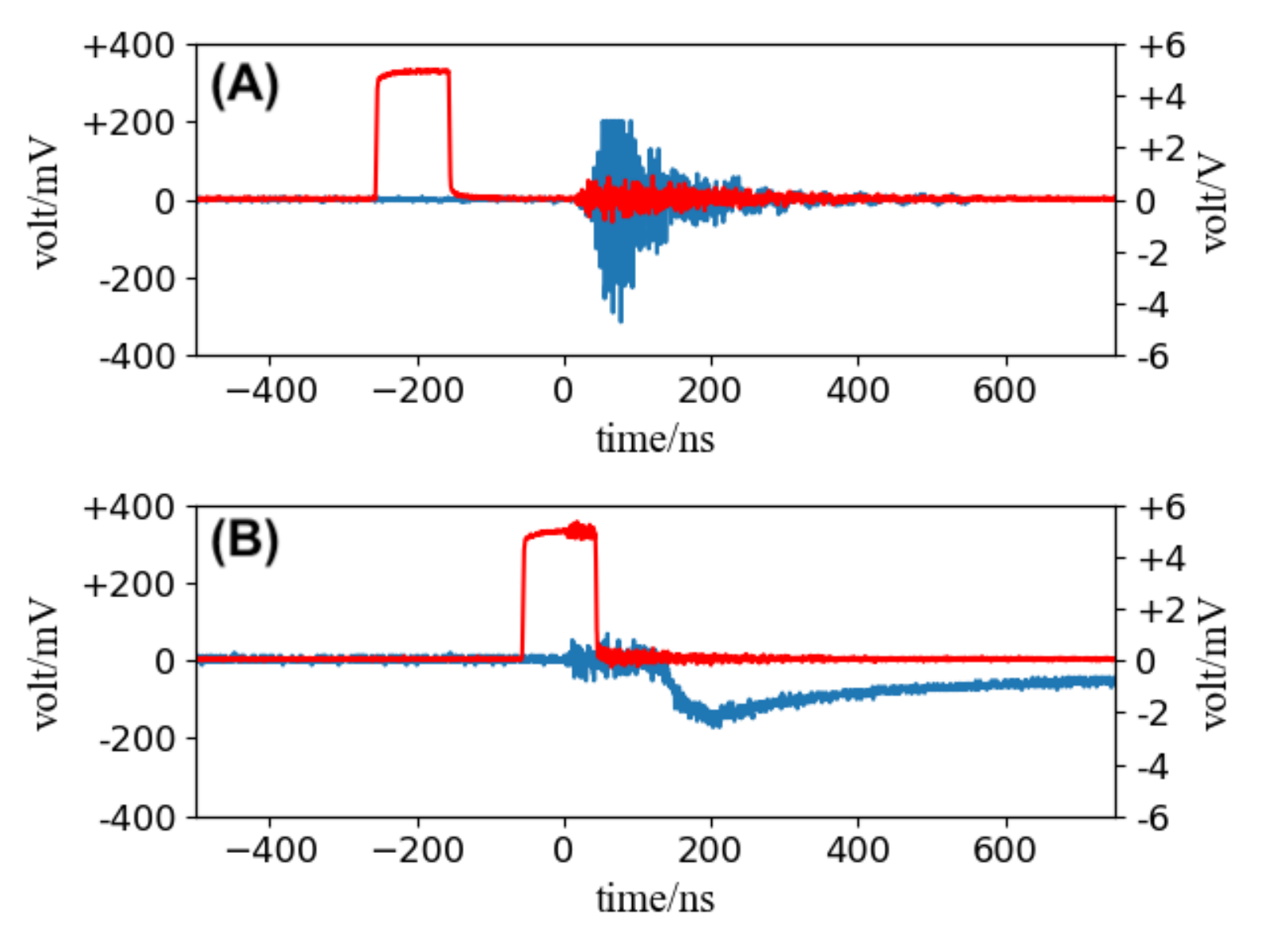} 
}
\subfigure{
\includegraphics[width=8.5cm]{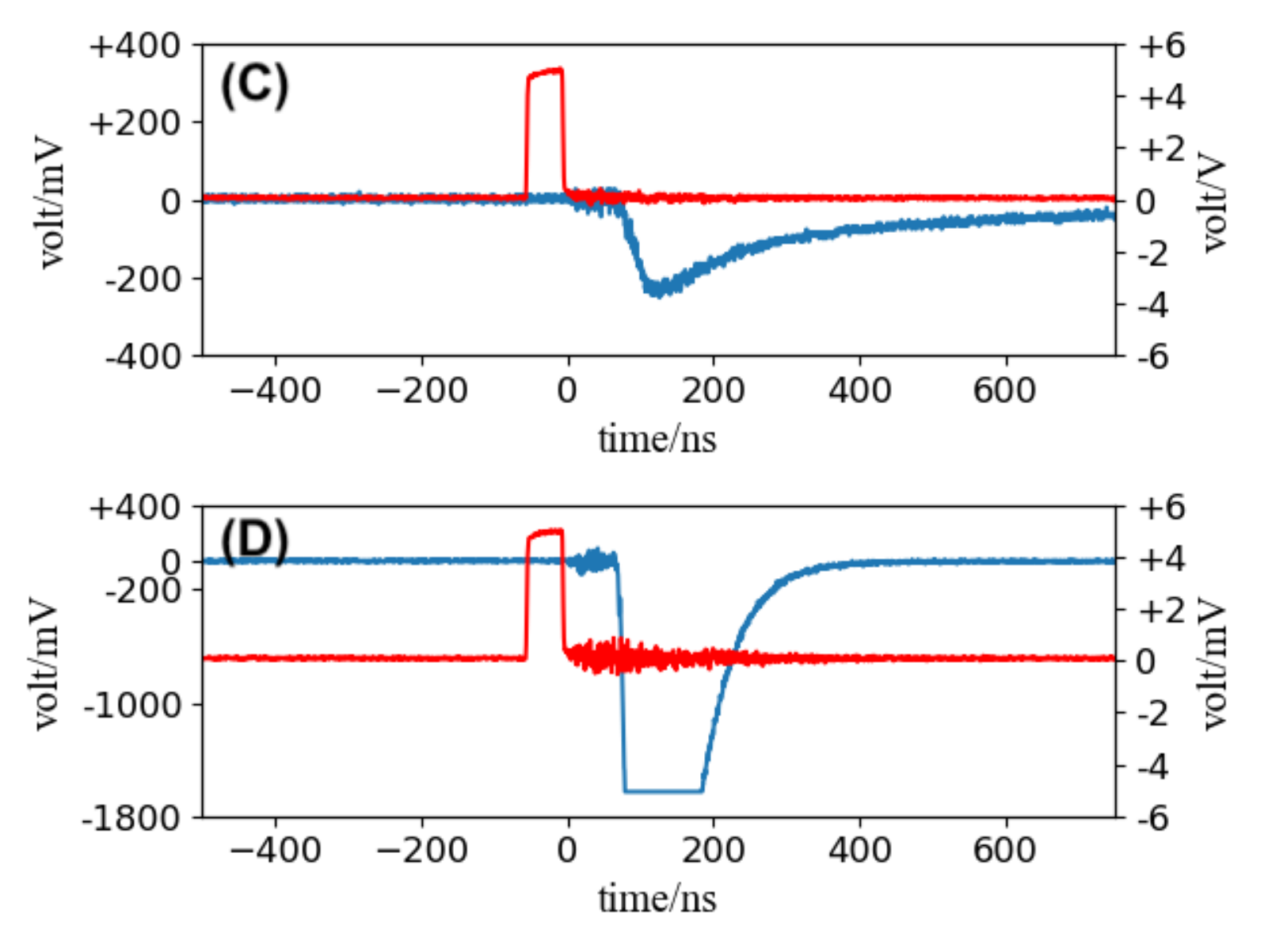} 
}

\caption{(Color online) Online test results at XG-III laser facility.
The square waves (red) are the trigger for the gPMT, and blue curves signals from a gPMT.
The gPMT is in normal-on mode.
The delay time of the trigger was turned to be $-260$ ns for (A), and $-60$ ns for (B), (C), and (D).
The pulse was turned to be 100 ns for (A)\&(B), and 50 ns for (C)\&(D).
In (A), the gPMT's power was turned off, while in others, the power was on.
The gPMT's gain in (D) was turned to be about 10 times higher than that in (B)\&(C).
Look at the context for details.}
\label{fig.XG.results}
\end{figure}

The GFD has been tested on the Xing Guang III (XG-III) laser facility
located at the Science and Technology on Plasma Physics Laboratory,  
the Laser Fusion Research Center,  Sichuan Province, China.
The facility has three laser beams with different wavelengths and duration, 
fs beam (pulse width 26 fs; maximum energy 20 J; wavelength 800 nm), 
ps beam (0.5--10 ps; 370 J; 1053 nm),
and ns beam (1.1 ns; 570 J; and 527 nm).

In our test runs, only a ps beam was used.
The typical energy of the ps beam is about 100 J.
The laser beam bombarded a gas jet.
The GFD was set at the forward angle of about $30^\circ$ to the laser beam direction. 
A plastic scintillator with a thickness of about 3 mm was used.
A collimator with a diameter of 20 mm was aligned before the GFD.
Electrons and ions induced by the laser and the gas target interaction were then detected by the GFD,
and the signals were recorded by a 200 MHz oscilloscope. 
The oscilloscope was triggered by a signal synchronous with the laser beam.
The same trigger signal was also used as the gate for the GFD. 
By tuning the relative time between the oscilloscope starting to record and the laser hitting the target,
we could turn off/on the GFD.

The results are shown in Fig.\ref{fig.XG.results}.
(A)  represents ``totally'' turn off the GFD.
For ``totally'', we mean that the power supply for the GFD is unplugged,
while, in the gating model, the power supply is plugged, 
and the PMT may not be biased when responding to a gate signal,
while other electronics of the GFD were still working.
But in the ``totally turn off'' case, the electronics were not working. 
From Fig.\ref{fig.XG.results}(A), there has a large signal. 
Its positive and negative amplitudes are almost equal.
It should be mentioned that the signal's peak showing at about 50 ns in the spectrum is due to circuit responding.
The original EMPs were very narrow ($<1$ ns), and quick (about 12 ns after laser-target interaction).
In fact, the GFD was located only about 4 m from the target, 
and it took photons only 12 ns to travel from the target to there.

By tuning the delay, we gated out the EMPs, which are shown in Fig.\ref{fig.XG.results}(B).
One can see that there has about 70 ns delay between the gate signal and the measured signal. 
This is due to the gPMT response time is about 70 ns.
The measured signals correspond to energetic electrons and ions from the laser-induced plasmablasts. 
In Fig.\ref{fig.XG.results}(C), 
by tuning the width of the gate signal from 100 ns to 50 ns,
the EMP can still be suppressed. 
By pushing this way, one can make the GFD responding time earlier.
Of course, this is risky of damaging the PMT if pushing too much. 
Keeping the same gate signal, and increasing the gPMT gain 6 times larger, 
the resulting spectrum is shown in Fig.\ref{fig.XG.results}(D). 
The signal was saturated at about 70--195 ns.

Based on the online test results shown in Fig.\ref{fig.XG.results}, the prototype GFD  developed by us does work as expected.
In this prove-of-principle test, the signals were induced by the plasma, which was composed of electrons and ions. The particle can be identified in future improved setups. 
Such as, together with traditional particle-identification methods like $\Delta E$-$E$, $TOF$, m/q, etc., 
energy and time signals from the GFD will help to identify the type of particles.
For example, by replacing image plates which are currently used in Thomson spectrometer focus planes\cite{ThompsonSpec2019-CPC,TomSpc2020.RN99} in typical HIL experiments with GFDs,
one can have extra time information, as well as an improved energy signal compared with that from the image plate, 
and then obtain much better particle-identification capabilities. 
Besides changed particles, this GFD can also be used to detect neutrons in harsh environments with a relative neutron-sensitive scintillator. 

Furthermore, photons may be emitted after time zero from excited nuclei and atoms which have relatively long metastable (or isomer for nuclei) states. 
As shown in Fig.\ref{fig.gPMT.timing}, 
if these photons arrive at the GDF away from the main peak,   
they can also be detected. 
With the capability of working in strong EMPs, the GFD will benefit HIL experiments in measuring neutrons, photons, and charged particles induced there.

\section{summary}

A major obstacle in laser nuclear physics studies is 
how to distinguish weak nuclear reaction product signals from very strong EMP signals induced by  HILs.
To overcome this difficulty, a gated fiber detector for HIL applications has been developed.
By using reflective foil, fiber, and gated PMT, 
strong EMPs which cause dysfunction of electronics 
in HIL environments are avoided.
By numerical simulation, the parameters like NA are optimized.
An online test shows that this prototype GFD can suppress EMP signals efficiently, and can be used for HIL environments. 
The GFD can respond about 70 ns later after the laser shooting on the target,
which makes it a good TOF detector to detect massive particles induced by HILs, 
as well as delayed gamma from excited states of nuclei in HIL-target interactions.

\begin{acknowledgments}
We would like to acknowledge the XG-III staff for operating the laser facility.
\end{acknowledgments}

%\bibliographystyle{apsrev4-1}
%\bibliography{ref}

\end{document}